\PassOptionsToPackage{table,xcdraw}{xcolor}
\documentclass[sigconf]{acmart}

\AtBeginDocument{%
  }

\usepackage[normalem]{ulem}
\usepackage{macros}
\usepackage{multirow}
\usepackage{adjustbox}

\usepackage{xurl}
\usepackage{tcolorbox}
\usepackage[]{mdframed}

\usepackage{algorithm}
\usepackage[noend]{algpseudocode}

\usepackage{enumitem}

\usepackage{array}
\newcolumntype{R}[2]{%
    >{\adjustbox{angle=#1,lap=\width-(#2)}\bgroup}%
    l%
    <{\egroup}%
}

\usepackage{xspace}

\usepackage{cleveref}
\usepackage{hhline}

\usepackage{graphicx}
\usepackage{subcaption}

\usepackage{pifont}
\newcommand{\cmark}{\ding{51}}%
\newcommand{\xmark}{\ding{55}}%

\usepackage[table]{xcolor}
\usepackage{soul}

\usepackage{booktabs} 

\begin{document}


\newcommand*{\name}{{\textsc{OpenTracer}}\xspace}

\newcommand*{\namea}{{\textsc{Trace2Inv}}\xspace}


\title{{\name}: A Dynamic Transaction Trace Analyzer for Smart Contract Invariant Generation and Beyond}

\author{Zhiyang Chen}
\orcid{0000-0002-2315-397X}
\affiliation{%
  \institution{University of Toronto}
  \city{Toronto}
  \country{Canada}
}
\email{zhiychen@cs.toronto.edu}

\author{Ye Liu}
\orcid{0000-0001-6709-3721}
\affiliation{%
  \institution{Singapore Management University}
  \city{Singapore}
  \country{Singapore}
}
\email{yeliu@smu.edu.sg}

\author{Sidi Mohamed Beillahi}
\orcid{0000-0001-6526-9295}
\affiliation{%
  \institution{University of Toronto}
  \city{Toronto}
  \country{Canada}
}
\email{sm.beillahi@utoronto.ca}

\author{Yi Li}
\orcid{0000-0003-4562-8208}
\affiliation{%
  \institution{Nanyang Technological University}
  \city{Singapore}
  \country{Singapore}
}
\email{yi_li@ntu.edu.sg}

\author{Fan Long}
\orcid{0000-0001-7973-1188}
\affiliation{%
  \institution{University of Toronto}
  \city{Toronto}
  \country{Canada}
}
\email{fanl@cs.toronto.edu}


\begin{abstract}
  Smart contracts, self-executing programs on the blockchain, facilitate reliable value exchanges without centralized oversight. Despite the recent focus on dynamic analysis of their transaction histories in both industry and academia, no open-source tool currently offers comprehensive tracking of complete transaction information to extract user-desired data such as invariant-related data. 
  This paper introduces {\name}, designed to address this gap. {\name} guarantees comprehensive tracking of every execution step, providing complete transaction information. 
  {\name} has been employed to analyze 350,800 Ethereum transactions, successfully inferring 23 different types of invariant from predefined templates. The tool is fully open-sourced, serving as a valuable resource for developers and researchers aiming to study transaction behaviors or extract and validate new invariants from transaction traces. 
  The source code of {\name} is available at \url{https://github.com/jeffchen006/OpenTracer}.
  \end{abstract}





\maketitle

\section{Introduction}

Blockchain technology has revolutionized the concept of decentralization on a global scale. One of its most impactful applications is smart contracts—computer programs that run atop blockchains to manage substantial financial assets and automate the execution of agreements among multiple trustless parties. The transaction histories of these smart contracts capture all execution data since their deployment, providing a rich source of information for analysis, such as smart contract invariants and user behavior patterns. 

\name is a dynamic analysis tool designed to parse raw transaction traces into human-readable formats, enabling deep analysis of specific trace segments to extract critical data, such as those related to invariants. This tool has been used in the recently accepted work, \namea~\cite{chen2024demystifying}, at The ACM International Conference on the Foundations of Software Engineering (FSE) 2024. In \namea, \name was utilized to analyze $350,800$ Ethereum transactions, from which it extracted pertinent data for $23$ different types of invariant templates across $8$ categories. It operates by parsing transaction trace data from an Ethereum archive node, subsequently locating and applying dynamic taint analysis and data flow analysis on specific trace snippets to extract the information desired by users. Throughout the development of \name, we have observed its capability to extract any detail from execution traces, establishing it as an invaluable tool for researchers aiming to generate, validate, and explore new invariants in future studies.

Existing tools related to \name are typically either closed-source or are designed to extract only essential information from transaction traces, tailored to meet specific, predefined requirements. These tools will be discussed in detail in Section~\ref{sec:related}. In contrast, \name is fully open-source and offers extensive customization options, enabling researchers to extract any information from transaction traces. This flexibility makes \name an exceptionally versatile tool suited to a wide range of research needs.
In this paper, we highlight three major applications of {\name}:
\begin{enumerate}
    \item As a stand-alone transaction explorer to generate function-level invocation trees of transactions with decoded storage accesses.
    \item As a dynamic analysis tool to perform deeper analysis (taint analysis/data flow analysis) on specific trace snippets to collect invariant-related data.
    \item As a replacement for modified archive Ethereum nodes in other tools.
\end{enumerate}

Through these use cases, we illustrate the versatility and potential of \name in advancing smart contract analysis and blockchain research.  

\section{Related Works}
\label{sec:related}
In this section, we review the existing tools and prior developments on transaction analysis, identifying the gaps that {\name} addresses.

\subsection{Transaction Explorers}
Numerous industry transaction explorers are currently available, offering free services that allow users to delve into transaction details on various blockchains, as detailed in Table~\ref{tab:txExplorers}. While these tools effectively provide basic transaction data at scale, they often fall short in offering comprehensive low-level transaction details, which are critical for program analysis. For instance, except for EthTx~\cite{ethtx}, all transaction explorers are closed-source. EthTx, though open-source, does not provide storage access information, thus limiting its functionality significantly. In contrast, {\name} is fully open-source and equipped to furnish a complete spectrum of transaction data, ensuring that users can extract and utilize the exact information they need from transactions. 

\begin{table}[!htbp]
    \caption{Comparison of Transaction Explorers: "Source" represents open-source availability. "Func." represents the display of function level invocation trees. "Store." represents storage access information visibility, including sload and sstore.}
    \label{tab:txExplorers}
    \begin{tabular}{|l|c|c|c|}
        \hline
    \textbf{Explorer} & \textbf{Open Source?} & \textbf{Func.} & \textbf{Store.}  \\ \hline
    Phalcon~\cite{phalcon}          & \xmark                            & \cmark & \xmark  \\ \hline
    EthTx~\cite{ethtx,ethtxsource}  & \cmark                            & \cmark & \xmark  \\ \hline
    AnyTx~\cite{anytx}              & \xmark                            & \cmark & \cmark  \\ \hline
    Tenderly~\cite{tenderly}        & \xmark                            & \cmark & \cmark \\ \hline
    OpenChain~\cite{openchain}      & \xmark                            & \cmark & \cmark  \\ \hline
    {\name}                         & \cmark                            & \cmark & \cmark  \\ \hline
    \end{tabular}
\end{table}

\subsection{Dynamic Smart Contract Invariant Generation}

Several notable works in the field focus on the generation of invariants from smart contract transaction histories.  SPCon~\cite{liu2022finding}  reconstructs likely access control models by analyzing function callers in historical transactions, while InvCon~\cite{liu2022invcon} and its followup work, InvCon+~\cite{liu2024automated}, employ pre/post conditions to derive invariants specifically designed to counteract prevalent vulnerabilities in smart contracts. However, these methods typically extract only essential information from the transaction logs. By employing {\name}, our approach can enhance the implementation of these existing methodologies by extracting more comprehensive data from the transactions, thus potentially increasing the robustness and applicability of the generated invariants. Additionally, there exists some other research dedicated to inferring invariants directly from the source code of smart contracts; while relevant, these approaches are not directly aligned with the focus of our work, which centers on dynamic analysis through transaction history.

\subsection{Transaction Anomaly Detection}
Significant efforts have been devoted to detecting anomalous transactions in blockchain environments, with key contributions such as TxSpector~\cite{zhang2020txspector}, Time-Travel Investigation~\cite{wu2022time}, Sereum~\cite{rodler2018sereum}, SODA~\cite{chen2020soda}, and The Eye of Horus~\cite{ferreira2021eye}. All of these approaches, except for The Eye of Horus, require modifications to the Geth client to access necessary execution traces, leading users to adopt a non-standard version of Geth, which must be updated with each new Geth release. 
Conversely, the Geth client offers a built-in debug RPC method through the \texttt{debug\_traceTransaction} RPC call, which allows for the replaying of transactions to retrieve execution traces. This feature is embedded in all Geth versions and stands as a vital RPC endpoint. Utilizing this RPC method can avoid the need for a modified Geth client. However, developers are required to parse the returned results, often requiring more development effort than modifying the Geth node itself. 
Notably, The Eye of Horus also utilizes the \texttt{debug\_traceTransaction} RPC call but lacks in providing a versatile tool for general transaction analysis. In contrast, {\name} not only harnesses this functionality to offer a general-purpose transaction analysis tool but also can possibly replace modified archive nodes for other applications, as demonstrated in Section~\ref{sec:evaluation}, where {\name} serves as an alternative to archive nodes for TxSpector~\cite{zhang2020txspector}.

\section{{\name} Overview}

\begin{figure*}[!htbp]
    \centering
    \includegraphics[width=\textwidth]{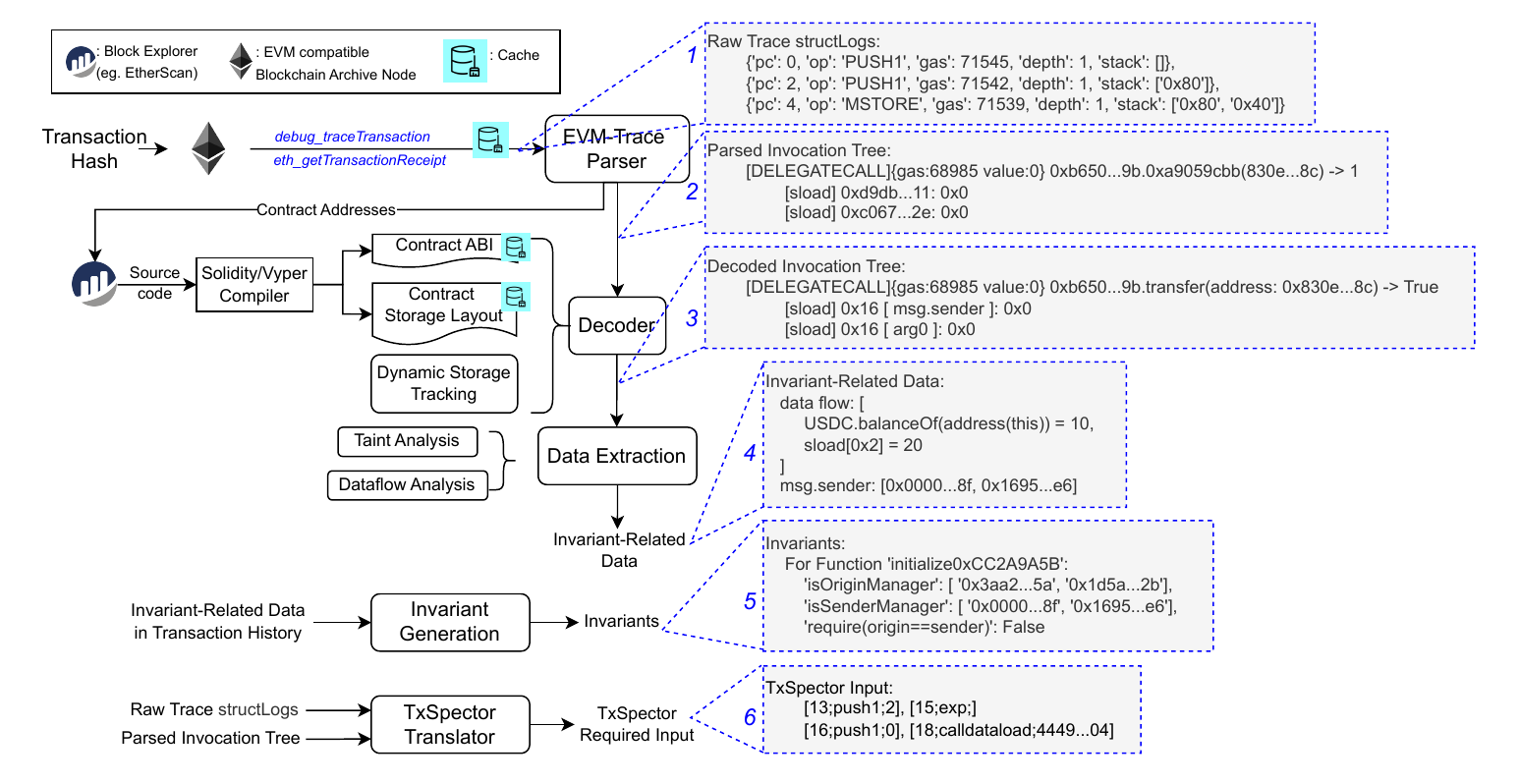}
    \caption{An Overview of {\name}}
    \label{fig:overview}
\end{figure*}

{\name} is a robust tool designed to meticulously capture and analyze transaction traces. As shown in Figure~\ref{fig:overview}, the process begins when {\name} receives a transaction hash. The following steps are executed:
(1) {\name} downloads the raw trace data from an Ethereum Archive Node.
(2) parses this data to construct an invocation tree,
(3) decodes function names, arguments, and return data utilizing function ABIs, and decodes storage accesses through dynamic storage tracking,
(4) finally analyzes specific trace snippets to extract user-desired data such as invariant-related data. 

\vspace{-3mm}
\subsection{Download and Augment Transaction Traces}
\name utilizes the Ethereum archive node RPC method \texttt{debug\allowbreak\_traceTransaction} to download transaction traces. These traces, referred to as \emph{structLogs} and highlighted in \textcolor{blue}{Box 1} of Figure~\ref{fig:overview}, consist of a series of executed EVM instructions. Each instruction is detailed in tuple format, encapsulating the program counter, opcode, gas metrics, and the current states of the EVM stack and possibly memory. Additionally, \name retrieves the transaction receipt using the \texttt{eth\allowbreak\_getTransactionReceipt} method, which supplements the trace with vital transaction details such as the block number and origin address. These RPC methods are supported by several providers such as Alchemy~\cite{alchemy}, Infura~\cite{infura}, and QuickNode~\cite{quicknode}, allowing users to use \name without maintaining a local Ethereum archive node, which is resource-intensive and requires substantial storage space.

\vspace{-3mm}
\subsection{EVM Trace Parser}
The parser module parses Ethereum Virtual Machine (EVM) trace data to construct an invocation tree. The parser identifies function entry and exit points using specific EVM opcodes categorized as "Function Enter" and "Function Exit".
The "Function Enter" opcodes include \texttt{call}, \texttt{callcode}, \texttt{staticcall}, \texttt{delegatecall}, \texttt{create}, and \texttt{create2}, marking the start of a function invocation. Conversely, "Function Exit" opcodes such as \texttt{stop}, \texttt{return}, \texttt{revert}, \texttt{selfdestruct}, an invalid opcode and scenarios where execution halts due to running "out of gas" signify the end of a function call.
This classification helps in constructing an invocation tree where each node is a detailed tuple recording the contract address, function selector, raw call data and return data. The parser also logs raw \texttt{sload} and \texttt{sstore} operations, setting the stage for more sophisticated decoding in subsequent steps.  An example of the output from this process is depicted in \textcolor{blue}{Box 2} of Figure~\ref{fig:overview}.

\vspace{-3mm}
\subsection{Decoder}
The decoder module enhances the interpretability of transaction trace data by processing function call data and storage accesses within the invocation tree. A sample output from this module is illustrated in \textcolor{blue}{Box 3} of Figure~\ref{fig:overview}, where both raw call data, raw return values, and storage accesses are decoded to more accessible formats.

\subsubsection{Decoding Function Call Data and Return Data}
Each node within the invocation tree is processed by the decoder, which accesses the contract's source code and ABI from blockchain explorers such as EtherScan~\cite{Etherscan}. The source code is compiled using Solidity~\cite{solidity} or Vyper~\cite{vyper} compilers to verify function signatures. This allows the decoder to translate the raw call and return data into a format that is not only human-readable but also accurately reflects the verified function signatures.

\subsubsection{Decoding Storage Accesses with Dynamic Storage Tracking}
This sub-module specializes in decoding storage keys and values captured in the EVM trace. By utilizing a dynamic storage tracking mechanism, \name is able to interpret complex storage structures. It tracks the computations of storage slots through EVM operations like \emph{sha3}, facilitating the decoding of storage accesses in the trace data. 
This capability enables a deeper understanding of how data is loaded and stored during contract execution. As shown in \textcolor{blue}{Box 3} of Figure~\ref{fig:overview}, \name tracks how a storage slot computed for each \texttt{sload} and \texttt{sstore} operation, providing a detailed decoded storage access information.

\vspace{-3mm}
\subsection{Data Extraction}
The data extraction functionality of {\name} identifies and analyzes specific trace snippets of interest, such as those accessing a particular storage slot or invoking a specific function. This module supports advanced techniques like taint analysis to trace malicious data flows and data flow analysis to monitor changes across stack, memory, and storage resulting from each EVM instruction. These capabilities are essential for pinpointing potential security vulnerabilities and understanding intricate data flows within smart contracts.
An example of invariant-related data extracted, demonstrating the module's output, is depicted in \textcolor{blue}{Box 4} of Figure~\ref{fig:overview}. This step ensures the comprehensive collection of data specified by users, effectively meeting diverse analytical needs.

\vspace{-3mm}
\subsection{Implementation and Optimization}
{\name} is developed in Python and consists of $12,049$ lines of code. The architecture allows for several optimizations including caching of intermediate results like raw trace data and transaction receipts as illustrated in Figure~\ref{fig:overview}. The system's design also supports future optimization of parallel processing of tasks such as data downloading, trace parsing, and decoding, significantly enhancing performance and efficiency.

\vspace{-3mm}
\subsection{Use case: Invariant Generation}
A fundamental application of {\name} is invariant generation for a given contract. Given a contract address as an input, users first obtain its transaction history using tools like TrueBlocks~\cite{TrueBlocks_Team_TrueBlocks_Lightweight_indexing_2022}. Then 
users may either employ existing methodologies or develop other techniques for data collection. 
Subsequently, {\name} is deployed to extract this data from each transaction. This extracted data is then leveraged to generate invariants based on predefined templates, concretizing any undetermined parameters within these templates. When \name was used in \namea, it already has $23$ different types of invariant templates across $8$ categories available for users to choose from.
An example of this functionality is illustrated in \textcolor{blue}{Box 5} of Figure~\ref{fig:overview}, showcasing a concrete invariant generated from the data extracted for a specific function. 

\vspace{-3mm}
\subsection{Use case: Trace Translation}
Another key application of {\name} is its ability to translate EVM trace data into formats required by other tools, enhancing its utility within the blockchain ecosystem.
This capability is pivotal as {\name} extracts comprehensive details like function call data, storage accesses, and data flow information for each EVM instruction. Users have the flexibility to devise their own translation methods in the parser or decoder modules, adapting the EVM trace data into the specific formats needed by other tools. For instance, the EVM trace data can be reformatted to meet the requirements of tools like TxSpector~\cite{zhang2020txspector}, effectively substituting the need for an archive node in such scenarios. An example of this translated trace is depicted in \textcolor{blue}{Box 6} of Figure~\ref{fig:overview}, showcasing how translated trace data is prepared for use by other analytical tools.

\section{Evaluation}
\label{sec:evaluation}
In this section, we evaluate the applicability and performance of {\name} by applying it to real-world smart contracts. 

{\name}, utilized in \namea, efficiently generates $23$ invariants across $8$ distinct categories. These categories include Access Control, Time Lock, Gas Control, Oracle Slippage, Re-entrancy, and Money Flow, where {\name} derives invariants solely from function-level invocation trees. In the "Special Storage" category, {\name} employs the contract's storage layout to decode and extract storage data for invariant generation. The "Data Flow" category utilizes both taint and data flow analyses to extract data needed. 

As detailed in Table \ref{tab:invariants}, our evaluation encompassed the transaction histories of $42$ contracts, known as victim contracts, from their deployment until they got hacked, totaling $350,800$ transactions. Using 70\% of these transactions as a training set, {\name} successfully generated $659$ invariants, averaging $15.69$ invariants per contract. Our results demonstrate that for each victim contract, at least one invariant was effective in protecting against its exploit. Notably, the most effective invariant \emph{GasStartUpperBound} alone is capable of protecting $30$ out of $42$ contracts from their respective exploits. These results underscore the effectiveness of {\name} in generating invariants that can safeguard smart contracts from common vulnerabilities.

\begin{table}[!htbp]
    \centering
    \caption{Summary of Evaluated Contracts and Transactions}
    \vspace{-2mm}
    \label{tab:invariants}
    \begin{tabular}{|l|l|}
    \hline
    \textbf{Metric}                    & \textbf{Value} \\ \hline
    \# Contracts Applied               & 42             \\ \hline
    Total Transactions Analyzed        & 350,800        \\ \hline
    Average Invariants per Contract    & 15.69          \\ \hline
    \# Contracts Protected for all invariants            & 42 out of 42   \\ \hline
    \# Contracts Protected for the best invariant            & 30 out of 42   \\ \hline
    
    \end{tabular}
\end{table}

Table~\ref{tab:punk} showcases the performance of {\name} when analyzing the transaction history of the Punk\_1 contract, comprising $31$ transactions. This assessment was conducted on a MacBook Pro with an Apple M2 chip, equipped with 8 CPU cores and 8 GB of RAM.

{\name} required a total of $8.55$ seconds to parse all $31$ transactions. The data collection process for invariant-related information, which involves complex operations such as taint analysis and data flow tracking, generally takes longer. Specifically, the extraction of data for $21$ invariants—excluding 2 Oracle Slippage invariants not applicable to Punk\_1—averages $3.965$ seconds per transaction. The time to extract data ranges from a minimum of $1.125$ seconds to a maximum of $5.242$ seconds per transaction. These metrics highlight {\name}'s efficiency in data collection. Additionally, the data collection process is designed to be parallelizable, allowing for simultaneous analysis of multiple transactions, which can significantly enhance efficiency. The translation of all $31$ transactions to TxSpector input format is completed in a mere $7.46$ seconds, underscoring the swift processing capability of {\name}. These performance metrics clearly demonstrate that {\name} is not only fast and efficient but also capable of handling large-scale transaction histories effectively.

\begin{table}[!htbp]
    \centering
    \caption{Performance Metrics of {\name} for Punk\_1 Transactions}
    \vspace{-3mm}
    \label{tab:punk}
    \begin{tabular}{|l|l|}
    \hline
    \textbf{Task}                             & \textbf{Time (s)} \\ \hline
    Parse All Txs                    & 8.55                    \\ \hline
    Max Data Collection Time Per Tx           & 5.242                   \\ \hline
    Min Data Collection Time Per Tx           & 1.125                   \\ \hline
    Avg Data Collection Time Per Tx           & 3.965                   \\ \hline
    Infer Invariants                          & 16.58                   \\ \hline
    Translate All Txs to TxSpector Format             & 7.46                    \\ \hline
    \end{tabular}
    \vspace{-3mm}
\end{table}

\section{Conclusion}

Smart contracts facilitate reliable blockchain transactions without centralized oversight, yet comprehensive tools for dynamic analysis of their transaction histories have been lacking. {\name} addresses this gap by providing detailed tracking of every execution step. Available as an open-source tool, {\name} offers extensive resources for developers and researchers to explore and validate new invariants, showcasing its effectiveness and efficiency. Its robust performance and accessibility position {\name} as a pivotal resource in advancing the security and understanding of smart contracts.



\newpage

\bibliographystyle{ACM-Reference-Format}
\bibliography{main}



\end{document}